\documentclass[twocolumn,twoside,slac_two]{revtex4}
\usepackage{graphicx}
\usepackage{fancyhdr}
\usepackage{amsmath,amssymb,amsbsy,amsfonts}
\begin{document}

\title{Relativistic solutions to the problem of jets with time--dependent 
       injection velocities}

%

\author{Sergio Mendoza (\textsl{sergio@astroscu.unam.mx})}
\affiliation{Instituto de Astronom\'{\i}a, Universidad Nacional 
    Autonoma de Mexico, AP 70-264, Ciudad Universitaria,
    Distrito Federal CP 04510, Mexico}
\author{ Juan Carlos Hidalgo (c.hidalgo@qmul.ac.uk)}
\affiliation{Astronomy Unit, School of Mathematical Sciences, Queen Mary, 
            University of London}
%

\begin{abstract}
  We present a ballistic description of the propagation of the working
surface of a relativistic jet.  Using simple laws of conservation
of mass and linear momentum at the working surface, we obtain a full
description of the jet flow parametrised by the initial velocity and mass
injection. This analysis will soon be applied to particular cases of 
time-dependent injection of mass and velocity into the jet.
\end{abstract}

\maketitle

\thispagestyle{fancy}


\section{Introduction}

  The apparent superluminal knots observed along the relativistic jets
of quasars and microquasars are usually interpreted as shock waves moving
through the jet.  It is not perfectly understood what is the mechanism that
can generate internal working surfaces that move along an astrophysical
jet, but it is generally accepted that the formation of these shocks is
produced by  a variation on the ejection flow velocity of the 
jet material \citep[see for example][and references within]{raga90}. 

  In this work we present a full relativistic generalisation of the
non--relativistic one dimensional dynamical description of internal
working surfaces made by \citet{canto00} that can easily be applied to
the most energetic jets associated with quasars, microquasars and GRB's.

\section{Dynamics of relativistic working surfaces}

  To follow the evolution of the working surfaces, we consider a source
ejecting material in a preferred direction \( x \) with a velocity \(
u(\tau)  \) and a mass ejection rate \( \dot{m}(\tau) \), both dependent on
time \( \tau \).

\begin{figure}
\begin{center}
  \includegraphics[width=0.40\textwidth]{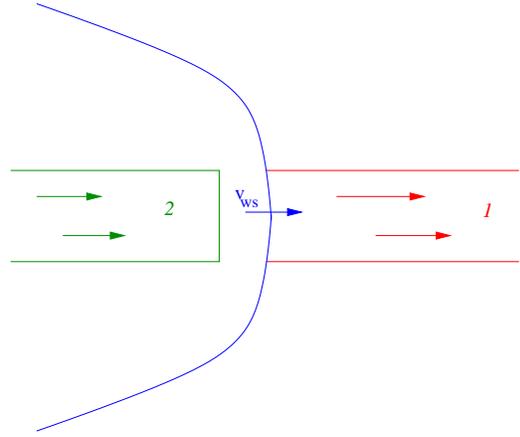}
\end{center}
  \caption{When a fast velocity flow 2  moves
     over a slow velocity flow 1, a working surface moving with velocity
     \( v_\text{ws} \) is generated as a result of the interaction.  }
\label{fig01}
\end{figure}

  Once the material is ejected from the source, we assume it will flow
in a free-stream fashion \citep[see e.g.][]{raga90}. The formation
of a working surface is studied as the intersection of two distinct
parcels of material ejected at times \( \tau_1 = 0 \) and \( \tau_2
= \Delta t \) labelled by their flow velocities \( u_1= u(\tau_1) \)
and \( u_2 = u(\tau_2) = u_1 + \alpha\,\Delta t \) respectively (see
Figure~\ref{fig01}). If \( \alpha > 0 \), the second parcel will eventually
reach the first parcel. At the time \( \tau_2 \), the distance between
the parcels is \( u_0 \Delta t \) and thus the time \( t_\text{m} \)
(measured in the reference frame of the source) when both parcels merge
is given by

\begin{equation} 
  \begin{split}
  t_\text{m} 
      =& \frac{1}{\alpha} u_1 \gamma^2 \left( u_1 \right) \left\{ 1 -
         \frac{u_1^2}{c^2} - \frac{ \alpha u_1 \Delta t }{ c^2 } \right\}, \\
      =& \frac{u_1}{\alpha} \left\{ 1 - \frac{ \gamma^2 \alpha u_1 \Delta
         t }{ c^2 } \right\},
  \end{split}
\label{eqn1.2}
\end{equation}

\noindent where \( \gamma^{-2}(u) :=  1 - u^2 / c^2 \) represents the Lorentz
factor of the flow with velocity \( u \).  The working surface is formed
at a distance \(  d_{f} = u_1 (t_m + \Delta t) \) from the source.

  Following the non--relativistic formalism first proposed by
\citet{canto00}, we assume that the working surface is thin and that
there are no mass losses within it (e.g. by sideways ejection of material
\citep[see][]{falle93,falle95}). Using the free-streaming condition,
we can then calculate the position \( x_\text{ws} \) of the working
surface from the downstream flow

\begin{equation} 
  x_\text{ws} = u_1 (t - \tau_1), 
\label{eqn1.3}
\end{equation}

\noindent or from the upstream flow

\begin{equation} 
  x_\text{ws} = u_2 (t - \tau_2). \label{eqn1.4}
\end{equation}

  On the other hand, since the flow is free--streaming, the velocity of
the working surface is given by the velocity \( v_\text{ws} \) of it's
centre of mass, which is determined by \citep{daufields}

\begin{equation} v_\text{ws} = \frac{ 1 }{ M_\gamma }
  \int_{\tau_1}^{\tau_2} \gamma\left(
    u(t) \right) \dot{m}(t) u(t)  \mathrm{d}t, \label{eqn1.5}
\end{equation}

\noindent where the weighted mass \( M_\gamma \) ejected between times 
\( \tau_1 \) and \( \tau_2 \) is

\begin{equation}
  M_{\gamma} = \int_{\tau_1}^{\tau_2} \gamma\left(u(t)\right) \dot{m}(t)
               \mathrm{d}t. 
\label{eqn1.6}
\end{equation}
       
  With this velocity, the position of the working surface is given by 
       
\begin{equation}
  x_\text{ws} = (t - \tau_2) v_{ws} + \frac{ 1 }{ M_\gamma } 
    \int_{ \tau_1 }^{ \tau_2 } \gamma\left( u(t) \right) \, \dot{m}(t) \, 
    u(t) \, (t - \tau_2) \, \mathrm{d}t.
\label{eqn1.7}
\end{equation}


  For a given value of position \( x_\text{ws} \), expressions
\eqref{eqn1.3}, \eqref{eqn1.4} and \eqref{eqn1.7} establish a relation
between the times \( \tau_1 \) and \( \tau_2 \).  The other is used to
eliminate t.  Taking \( \tau_2 \)  as a parameter, we can construct
the position and velocity of the working surface as a function of \(
\tau_2 \) and calculate relevant quantities such as the energy available
on the working surface.

  To calculate the amount of energy radiated as the working surface moves,
we take into account the energy \( E_0 \) the material had when it was
ejected, i.e. 

\begin{equation}
  E_{0} = \int_{ \tau_1 }^{ \tau_2 } \dot{m}(\tau) \, \gamma\left( u(\tau) 
    \right) \, c^2 \mathrm{d} \tau,
\end{equation}

\noindent and the energy \( E_\text{ws} \) of the material inside the working 
surface, which is given by 
      
\begin{equation}
  E_{ws} = m \gamma_{ws} c^2,
\end{equation}

\noindent where the Lorentz factor \( \gamma_\text{ws} \) of the working 
surface material is given by \( \gamma_\text{ws}^{-2} = 1 - v_\text{ws}^2/
c^2 \).  

  If we assume now that the energy loss along the jet, \( E_r = E_0 -
E_\text{ws} \), is completely radiated away, then the luminosity \( L =
\mathrm{d} E_\text{r} / \mathrm{d} t \) of 
the working surface given by


\begin{equation}
  \begin{split} 
  L &= \frac{ \dot{m}(\tau_2) c^2 }{  \mathrm{d} t / \mathrm{d} \tau_2 }
       \left\{ \gamma_\text{ws} + \frac{ m }{ M_\gamma } 
       \frac{ \gamma_\text{ws}^3 \gamma_2 }{ c^2 } \left( v_\text{ws} 
       u( \tau_2) -  v_\text{ws}^2 \right) - \gamma_2 \right\} -     \\
      & - \frac{ \dot{m}(\tau_1) c^2  }{ \mathrm{d} t / \mathrm{d} \tau_2 }
       \frac{ \mathrm{d} \tau_1 }{ \mathrm{d} \tau_2} \left\{ \gamma_\text{ws} +
       \frac{m}{ M_{\gamma} }  \frac{ \gamma_\text{ws}^3
       \gamma_1 }{ c^2 } \left( v_\text{ws} u( \tau_1) -  v_{ws}^2 \right) 
       - \gamma_1 \right\} ,
  \end{split}
\label{eq-lumino}
\end{equation}
   
\noindent where the Lorentz factors \( \gamma_{1,2}^{-2} :=
1 - v^2(\tau_{1,2}) / c^2  \) and we keep \( \tau_2 \) as a free parameter
in this expression.

\section{Conclusion}

  We have shown how a full relativistic solution can be constructed to the
problem of a ballistic working surface travelling along an astrophysical
jet.  Our main goal is to find analytic and
numerical solutions to equation~\eqref{eq-lumino} so that we can compare
with actual observations of high--energy jets.  This will be published
elsewhere soon.

\section{Acknowledgements}

  We would like to thank Jorge Cant\'o and Alex Raga for discussions
about their non--relativistic work.  We also appreciate the comments
made by William H. Lee, Xavier Hernandez and Enrico Ramirez--Ruiz about
the possible relativistic solutions to the problem.  JCH acknowledges
financial support granted by CONACyT~(179026). SM gratefully acknowledges
support from DGAPA (IN119203-3) at Universidad Nacional Aut\'onoma de
M\'exico (UNAM).

%

\bibliography{stan}





\end{document}